\documentclass[a4paper,11pt]{article}
\usepackage{pos}

\title{Reconstruction of sub-threshold events of cosmic-ray radio detectors using an autoencoder}
\ShortTitle{Autoencoder for cosmic-ray radio detectors}

\author*[a]{P.~Bezyazeekov}
\author[b]{D.~Shipilov}
\author[c]{D.~Kostunin}
\author[d]{I.~Plokhikh}
\author[d]{A.~Mikhaylenko}
\author[e]{P.~Turishcheva}
\author[f]{S.~Golovachev}
\author[f]{V.~Sotnikov}
\author[g]{E.~Sotnikova}

\affiliation[a]{Applied Physics Institute, Irkutsk State University, 664020 Irkutsk, Russia}
\affiliation[b]{X5 Retail Group, Moscow, 119049 Russia}
\affiliation[c]{DESY, 15738 Zeuthen, Germany}
\affiliation[d]{Novosibirsk State University, 630090 Novosibirsk, Russia}
\affiliation[e]{Innopolis University, 420500 Innopolis, Russia}
\affiliation[f]{JetBrains Research, 194100 St. Petersburg, Russia}
\affiliation[g]{Sobolev Institute of Mathematics, 630090 Novosibirsk, Russia}

\forColl{Tunka-Rex} 

\emailAdd{contact@tunkarex.info}

\abstract{
Radio detection of air showers produced by ultra-high energy cosmic rays is a cost-effective technique for the next generation of sparse arrays. The performance of this technique strongly depends on the environmental background, which has different constituents, namely anthropogenic radio frequency interference, synchrotron galactic radiation and others. These components have recognizable features, which can help for background suppression. A powerful method for handling this is the application of convolution neural networks with a specific architecture called autoencoder. By suppressing unwanted signatures, the autoencoder keeps the signal-like ones. We have successfully developed and trained an autoencoder, which is now applied to the data from Tunka-Rex. We show the procedures of the training and optimization of the network including benchmarks of different architectures. Using the autoencoder, we improved the standard analysis of Tunka-Rex in order to lower the threshold of the detection. This enables the reconstructing of sub-threshold events with energies lower than 0.1 EeV with satisfactory angular and energy resolutions.
}

\FullConference{37$^{\rm{th}}$ International Cosmic Ray Conference (ICRC 2021)\\
		July 12th -- 23rd, 2021\\
		Online -- Berlin, Germany}

\begin{document}
\maketitle

\section{Introduction}
Digital radio arrays became mature technique for the air-shower detection in last years~\cite{Schroder:2016hrv}.
The engineering arrays, e.g. Tunka-Rex, has shown the ability of energy spectrum~\cite{Tunka-Rex:2019msu} and mass composition reconstruction~\cite{Bezyazeekov:2018yjw} as well as was used for the independent cross-check of energy scale of Tunka-133 and KASCADE-Grande~\cite{Apel:2016gws}.
After decommission of Tunka-Rex in 2019 we focus on the open data~\cite{trvo} and development of techniques for the future arrays, e.g. aperture estimation and self-triggering for radio~\cite{Tunka-Rex:2018brd,lenokicrc2021}.
In this work we continue developing sub-threshold reconstruction using deep learning started in Ref.~\cite{Shipilov:2018wph}.

The efficiency of the method of reconstruction the primary particle parameters by measured radio emission, initiated with related air-shower, is sensitive to background characteristics.
High-level background may distorts the air-shower pulses up to impossibility of reconstruction in case of low amplitude pulses.
Standard methods of radio reconstruction uses the absolute amplitudes of air-shower pulses as main low-level characteristic of measured data.
Reconstruction of primary particle parameters is performed by analysis of lateral distribution of measured amplitudes.
For reconstruction the absolute amplitude of measured pulse in background conditions one can use analytical parametrization of mean background amplitude for reconstruction the amplitude of air-shower pulse, which works well in case of white noise and can be useful in some cases.
However, radio background in real conditions is non-white and have specific features.
Convolutional neural networks are effective tools for solving such type of tasks.

We show our approach to solve the background problem with autoencoder based on convolutional neural networks.
Autoencoder is tuned to extract background-related characteristics from the measured radio data, and remove them from the signal traces leaving only air-shower pulses.
We use the peak times of denoised signals and reconstruct the lateral distribution for further reconstruction of energy of primary particle.

\section{Autoencoder architecture}
Autoencoder is an architecture of artificial neural network, which codes the input data to abstract representation and after decodes it to as similar to input as possible.
It can be used for problems of dimensionality reduction, denoising, anomaly detection and others.
Structurally autoencoder consists of two parts: encoder and decoder.
The encoder consists of a set of layers with decreasing number of neurons at each next layer and produces compressed representation of original input or the code.
After encoding decoder starts to work.
In typical case, the decoder has the same structure as encoder but in opposite orientation.
The goal of decoder is to reconstruct original input from the code as close as possible.
As we have small-size layer in the center of the network, some part of the information about input data is inevitably lost.

Denoising autoencoder is specific architecture which is proposed to removing noise from the data.
During training of this type of autoencoder, it takes the noised data at input, and after processing compares output with the same data, but without noise.
This way it tunes to remove noise-related information and returns only useful data.

\begin{figure}
    \centering
    \includegraphics[scale=0.48]{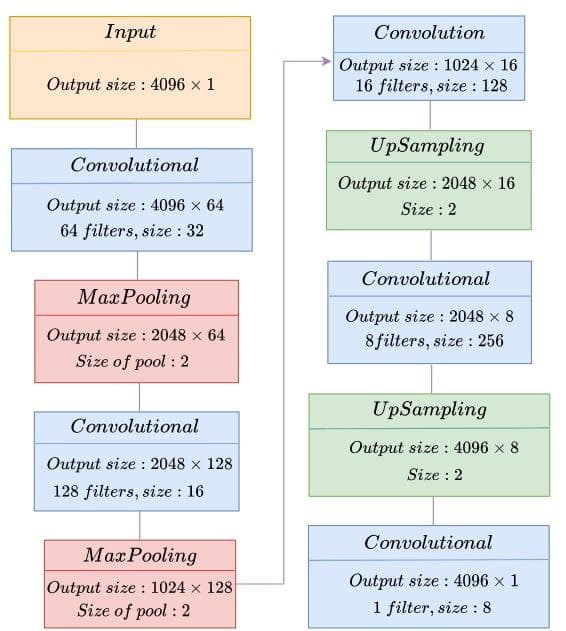}
    \caption{Autoencoder architecture}
    \label{fig:my_label}
\end{figure}

As we have data in one-dimensional format (signal trace of measured amplitude of electric field), for this study we use one-dimensional architecture of autoencoder.
For defining the optimal configuration for reconstruction the air-shower pulse we tested a number of autoencoder configurations with various number of layers and filters, size of filter, loss function etc.
For preliminary estimation of the efficiency of autoencoder we use binary cross entropy and visual analysis of performance results. 
Detailed view of chosen architecture is shown in Fig~\ref{fig:my_label}.
Loss function is ReLU~(rectified linear unit).
Training of AE was performed with Tensorflow+Keras on GPU server using uDocker virtualisation.

\section{Denoising of radio data with autoencoder}

For training the autoencoder we prepared the dataset consisting of 25000 CoREAS~\cite{Huege:2013vt} simulations based on CORSIKA software~\cite{HeckKnappCapdevielle1998} and library of background measured at Tunka-Rex consisting of 650000 samples.
Training dataset is formed by adding simulated air-shower pulses with background traces.
The air-shower pulse is located around the fixed position with random shift with range of $\pm$50 ns. Each sample contains one reconstructed polarization of measured/simulated electric field.
Before processing with autoencoder traces are upsampled with factor 16 (0.3125 nanoseconds per count instead of 5 nanoseconds per count in raw data).
Length of the trace is 4096 counts, which corresponds to 1.28 microseconds.
For correct performance of autoencoder we normalise the amplitude of each trace to [0:1] range.
Due to this we can not reconstruct the amplitude directly for very low signals (where signal-to-noise ratio close to unity) and develop a special procedure for reconstruction which is described in detail in next section.

\section{Reconstruction of real data}
In the frame of this work we aim at the detection of sub-threshold events not accessible with standard reconstruction.
We have selected events triggered by Tunka-133 and which have energy reconstructed by Tunka-133 in between $10^{16}$ and $10^{17}$\,eV.
Since the energy threshold of Tunka-Rex starts from about $10^{17}$\,eV this set has almost no overlap with standard Tunka-Rex reconstruction.
We decreased the internal threshold of autoencoder, which was set to value corresponding to 5\% probability of false positive for single signal.
Since we used normalization, the autoencoder returns amplitudes in the range of $[-0.5; 0.5]$ and we reduce threshold from $0.395$ to $0.2$.

\begin{figure}[t]
	\centering
	\includegraphics[height=0.47\linewidth]{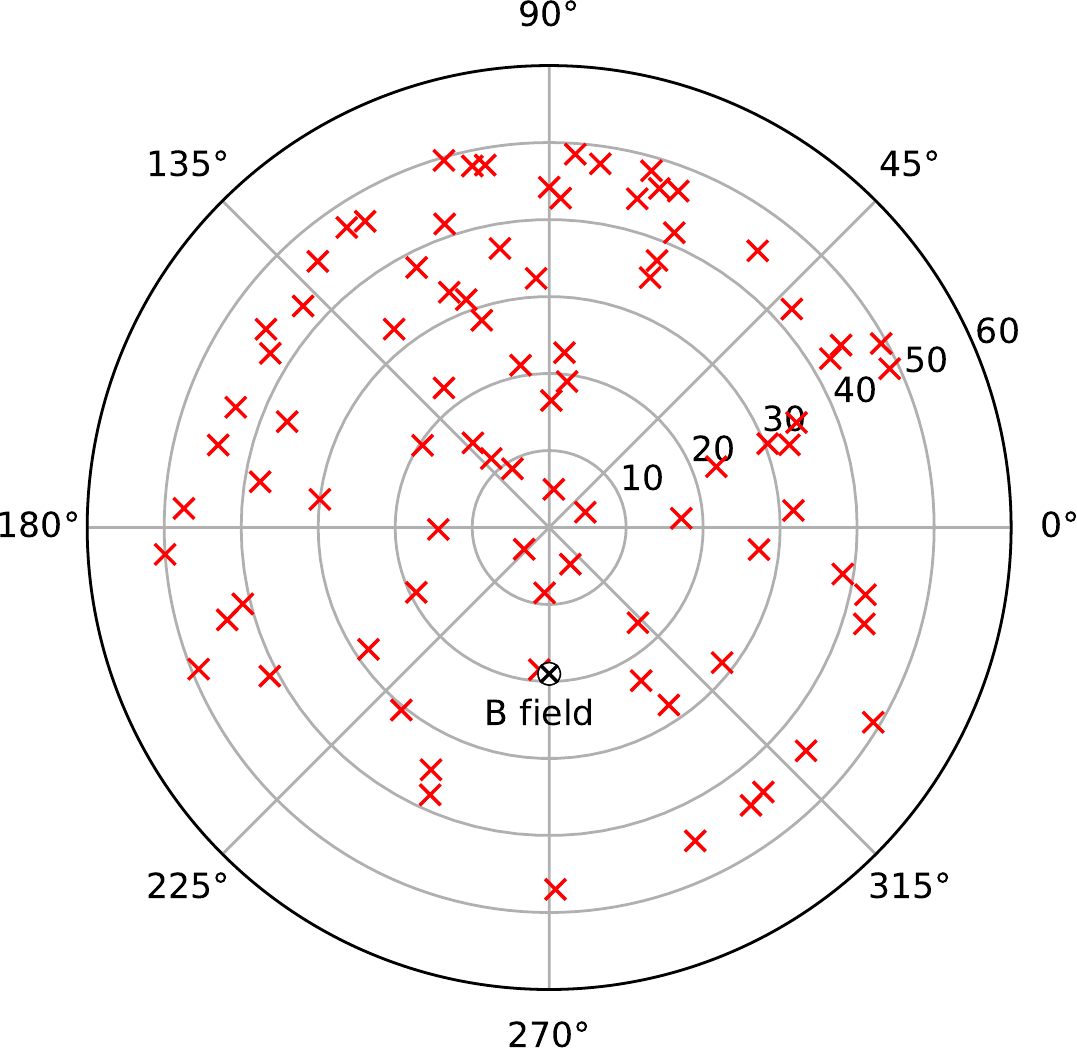}~~~~\includegraphics[height=0.47\linewidth]{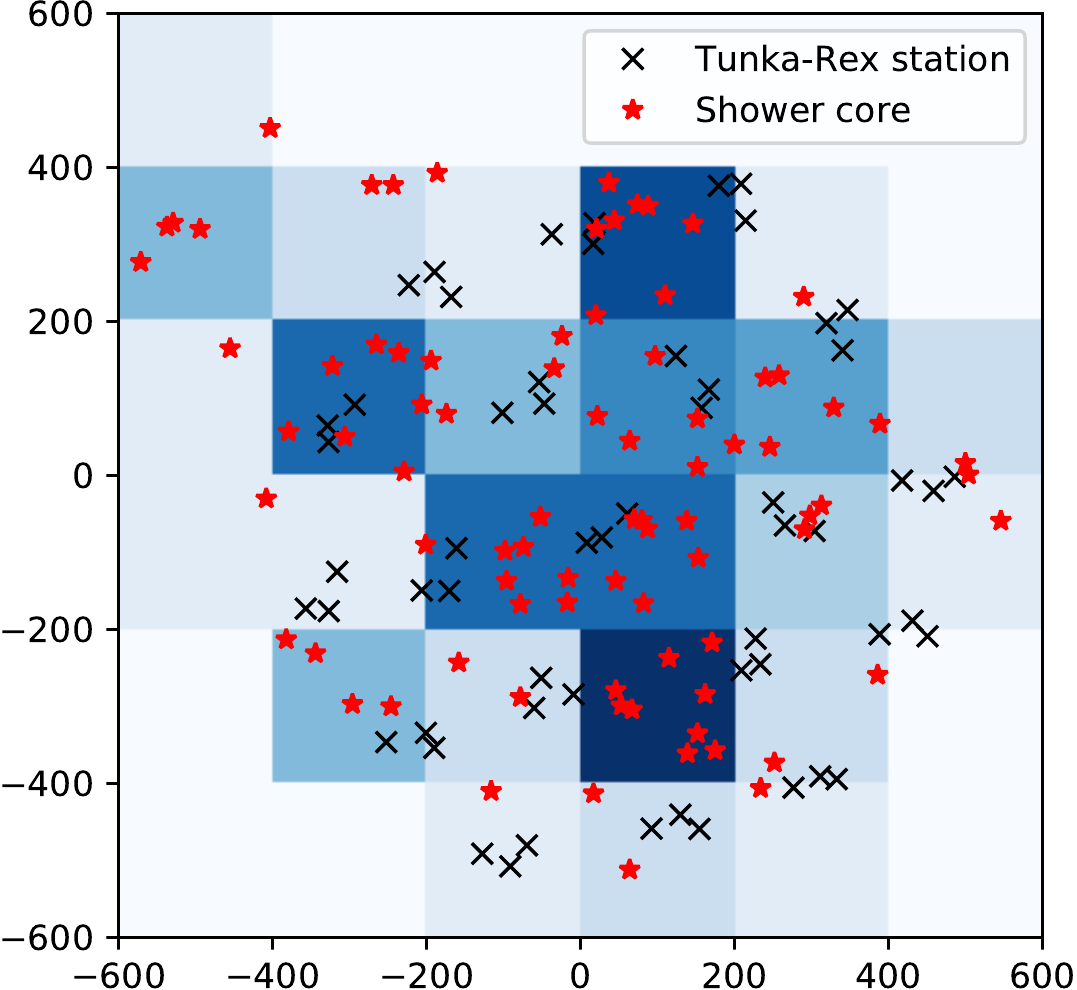}
	\caption{Distribution of events detected by autoencoder before quality cuts.
	\textit{Left:} arrival directions of detected events.
	One can see geomagnetic asymmetry, which indicated that the significant fraction of events are true positives.
	\textit{Right:} distribution of the cores of detected events taken from trigger, Tunka-133.
	No hotspots are obtained, which shows that autoencoder does not trigger on known local sources of the noise (e.g. central DAQ or power lines).}
	\label{fig:nocuts}
\end{figure}

Fig.~\ref{fig:nocuts} showes detected events passed standart quality cuts (difference between Tunka-133 and Tunka-Rex arrival directions less than $5^\circ$, antennas are clustered, etc.).
The distributions of arrival directions and core positions of detected events indicate significant fraction of true positives, and we introduce additional quality cuts and algorithms, which allows us not only to reject false positives, but also reconstruct primary energy.

\begin{figure}[t]
\centering
Original traces:
\includegraphics[width=0.5\linewidth]{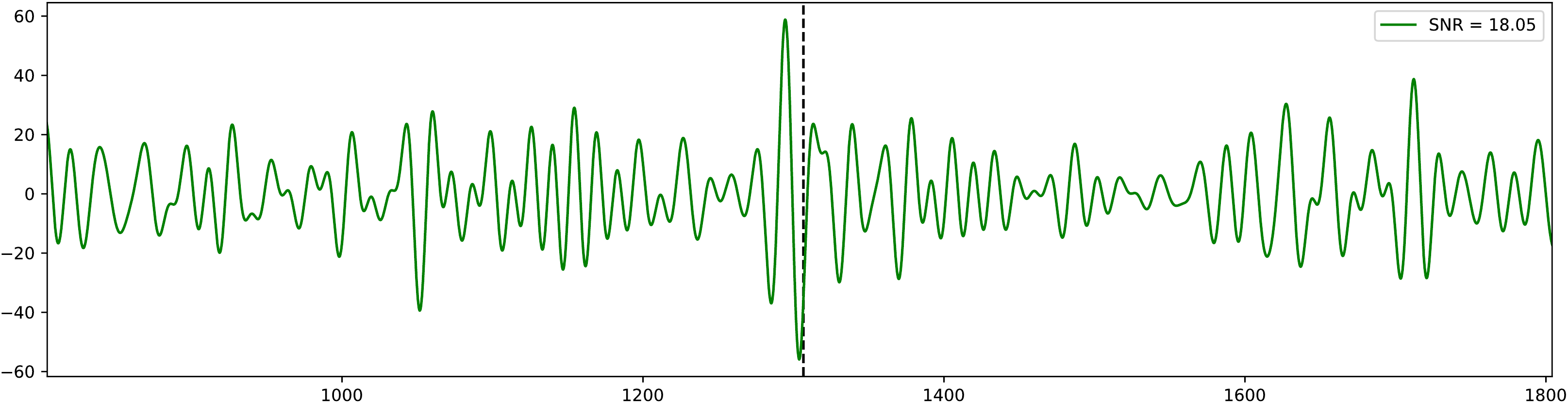}~~\includegraphics[width=0.5\linewidth]{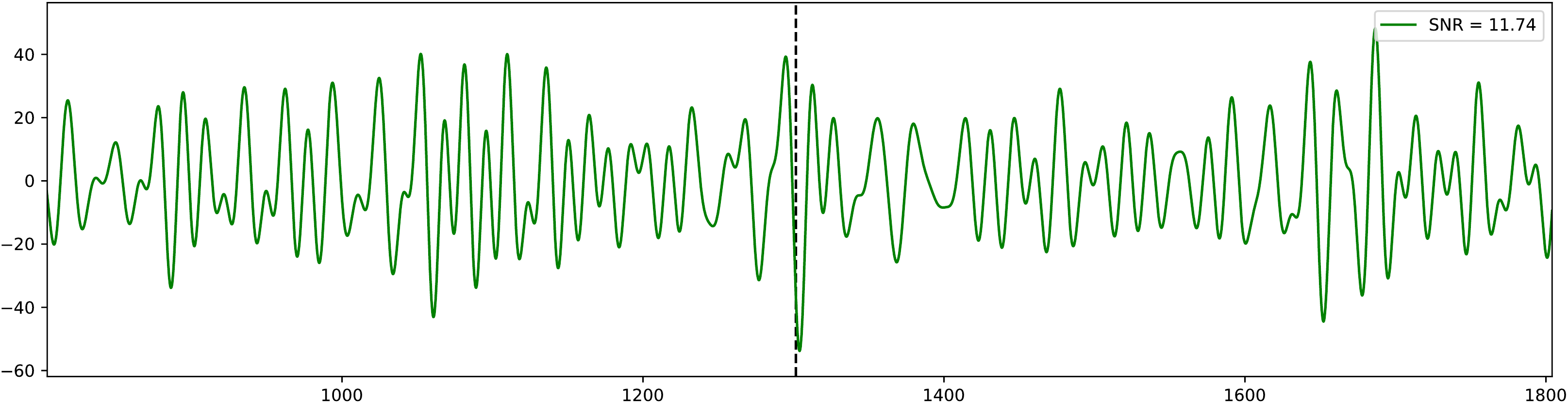}\\
\includegraphics[width=0.5\linewidth]{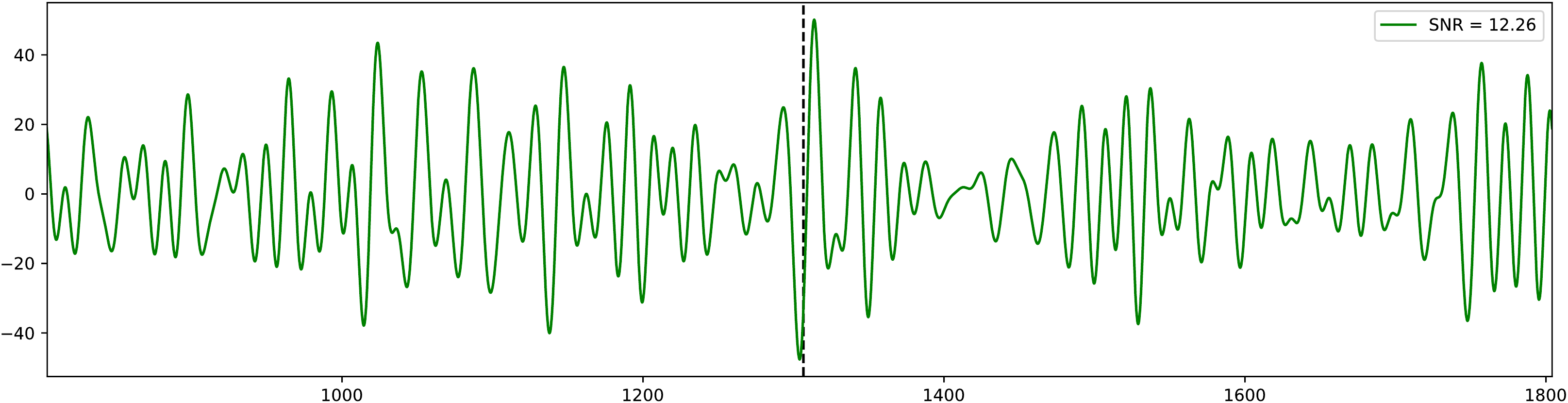}~~\includegraphics[width=0.5\linewidth]{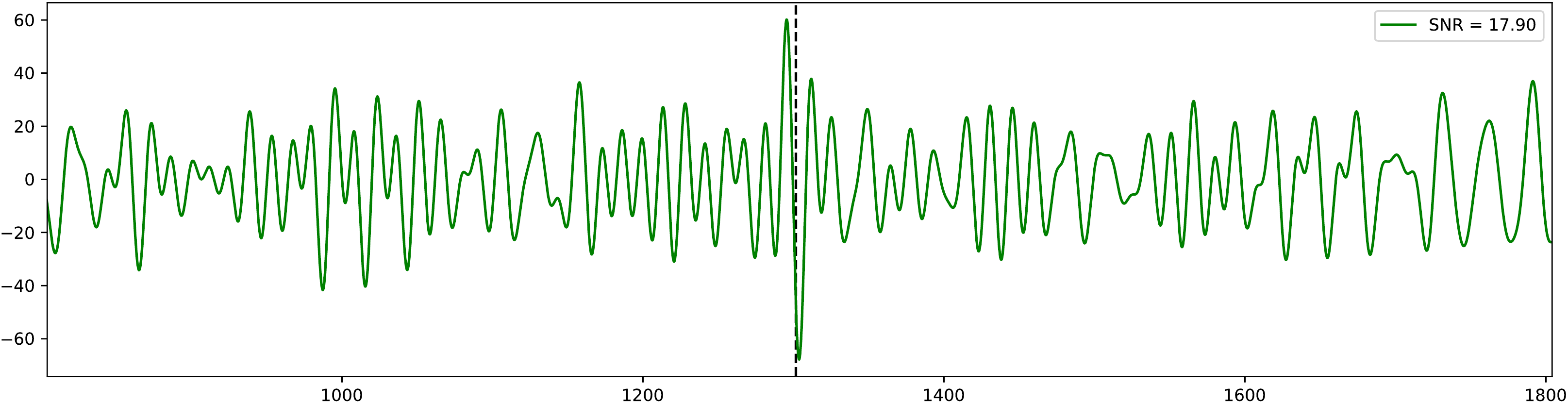}\\
\includegraphics[width=0.5\linewidth]{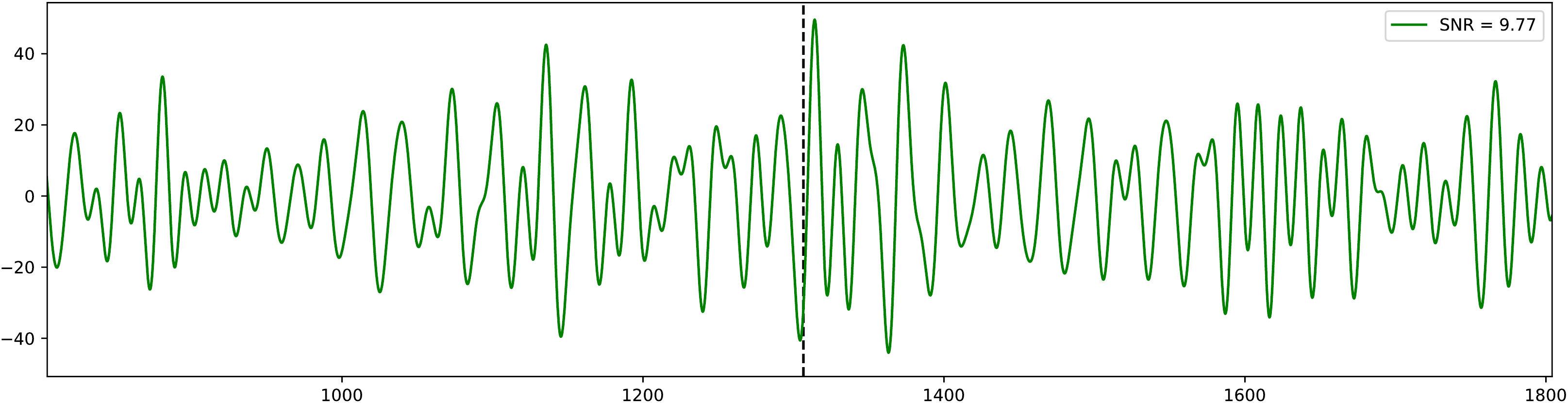}~~\includegraphics[width=0.5\linewidth]{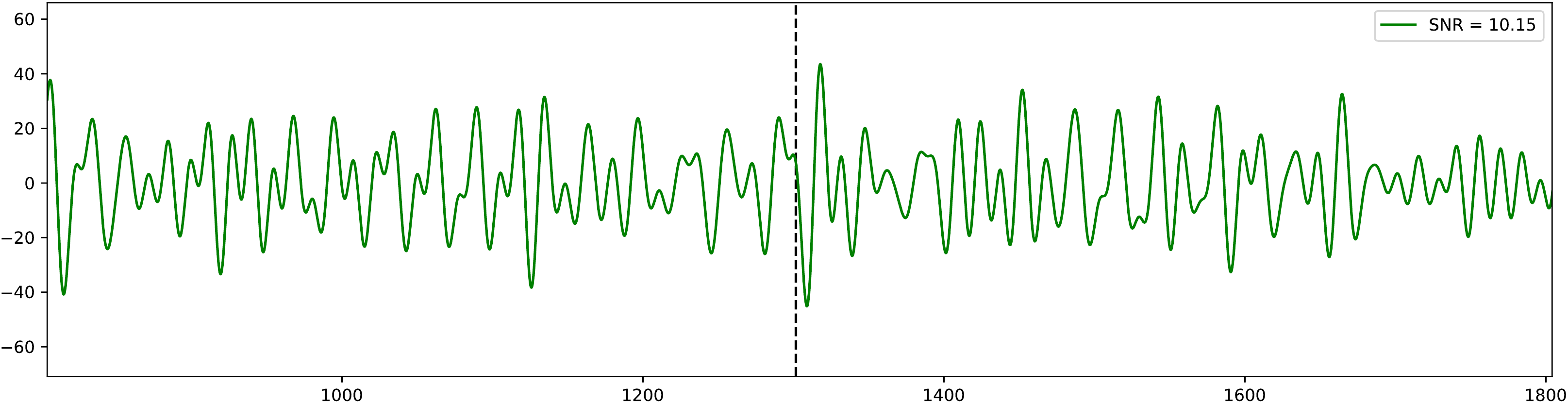}\\
Sum:\\
\includegraphics[width=0.5\linewidth]{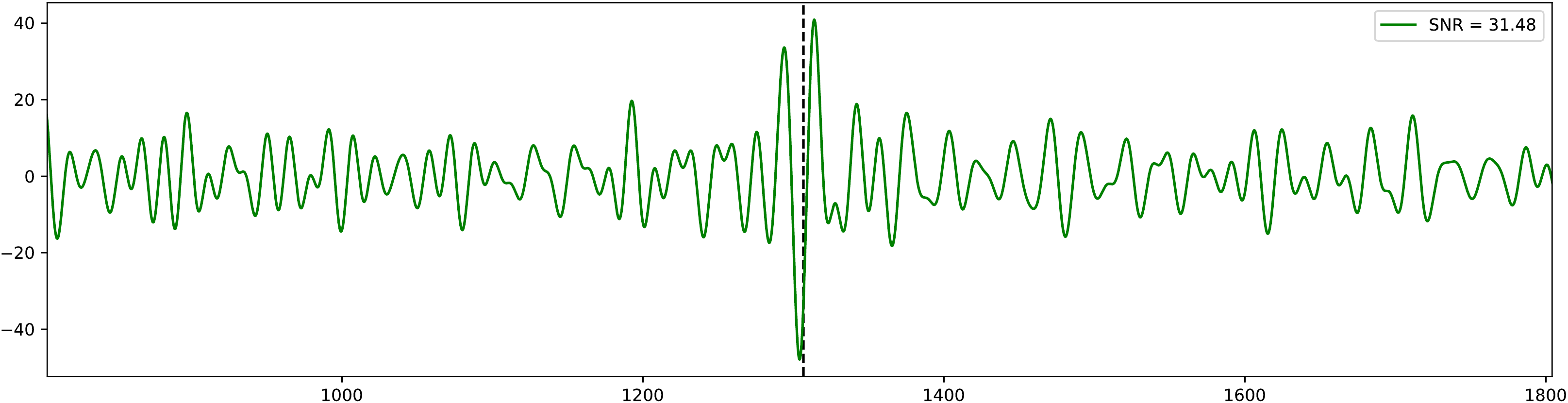}~~\includegraphics[width=0.5\linewidth]{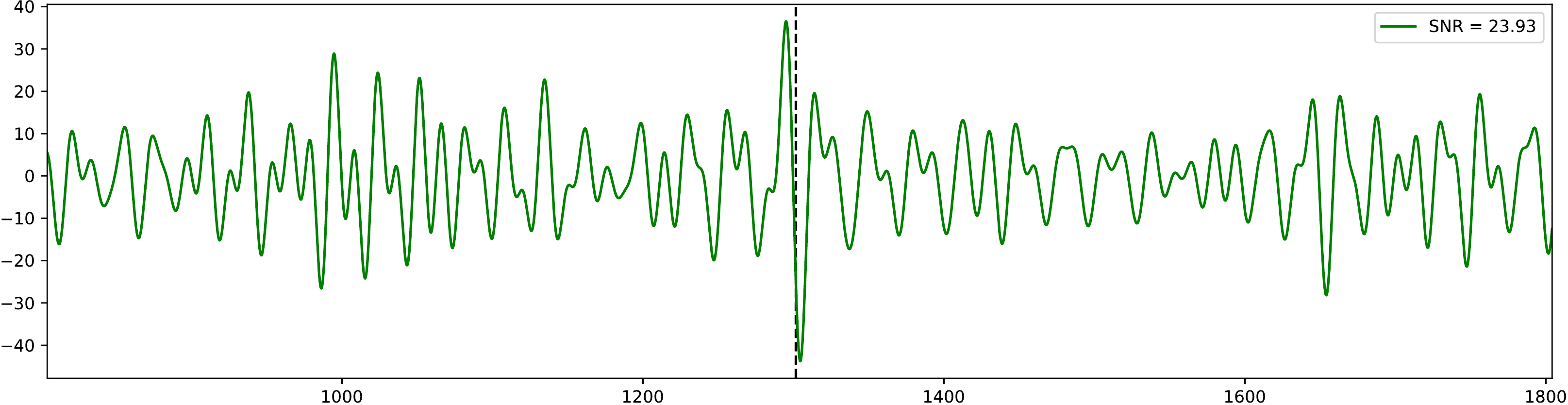}
\caption{Example of two events with $E=30$~PeV detected with autoencoder. Dashed lines indicate the reconstructed peak position.}
\label{fig:sum_traces}
\end{figure}

We developed so-called \textit{adaptive LDF}\footnote{Lateral distribution function (LDF) is a distribution of amplitudes projected on shower axis.} algorithm exploiting the idea of coherent production of radio emission during air-shower development (see Fig.~\ref{fig:sum_traces})
This algorithm sums radio traces using peak position reconstructed by autoencoder and synthesises effective amplitude at effective distance from the shower core.
Assuming uncorrelated background the signal-to-noise ration (SNR) of synthesised signal should have increase by factor of $\sqrt{N}$, where $N$ is a number of original signals.
This feature is used for detection of the false positives and exclusion them from the analysis, we add additional cut $\mathrm{SNR}>16$ for the synthesised signals.

Adaptive LDF does not depend on the time resolution of the detector and requires only position of the radio pulse in the trace.
Coherent sum features its maximal performance for the signals spatially close to each other, since they are produced in the same part of air-shower and their shapes are almost identical.

\begin{figure}[t]
\centering
\includegraphics[height=0.49\linewidth]{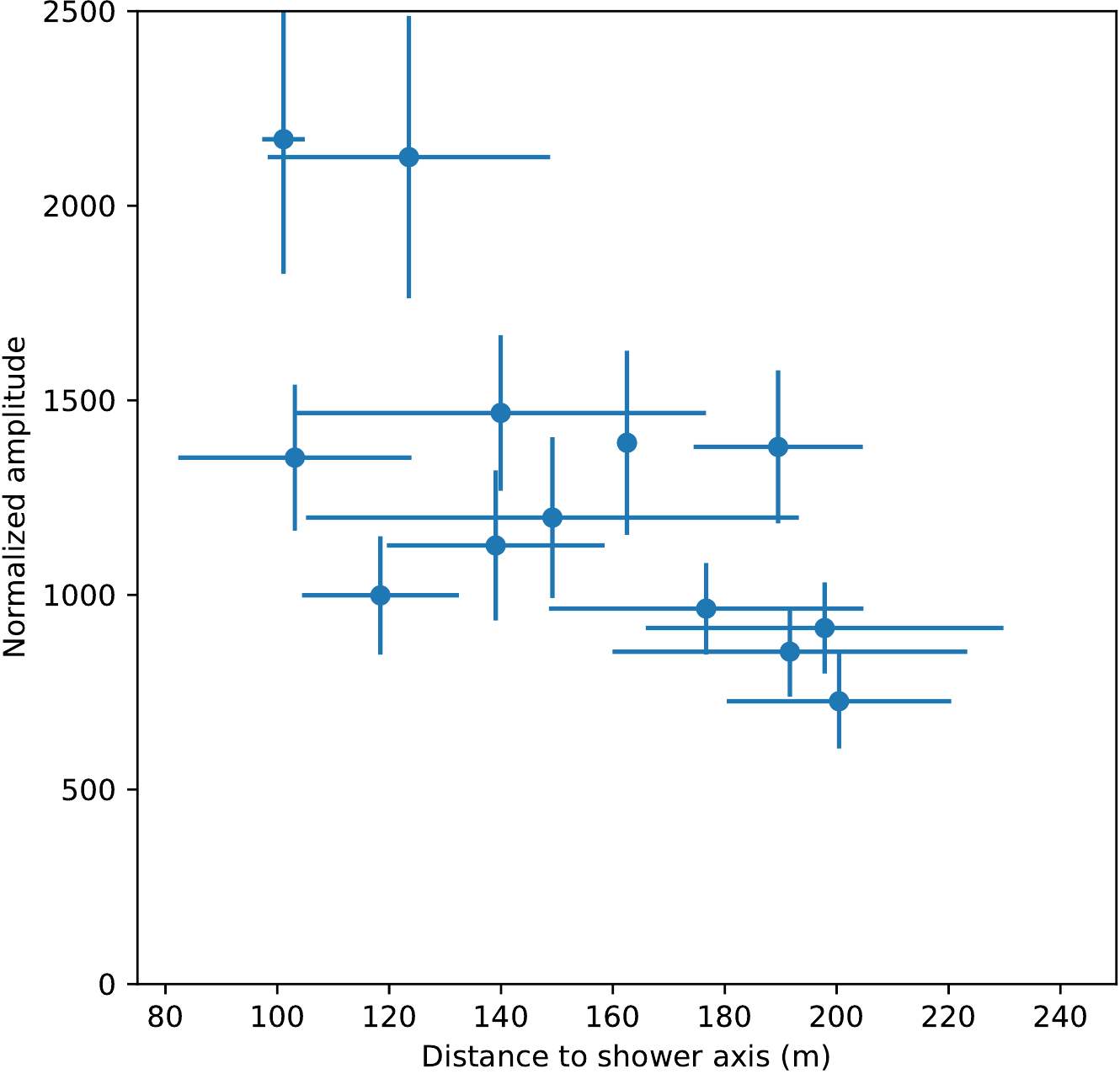}~~~\includegraphics[height=0.49\linewidth]{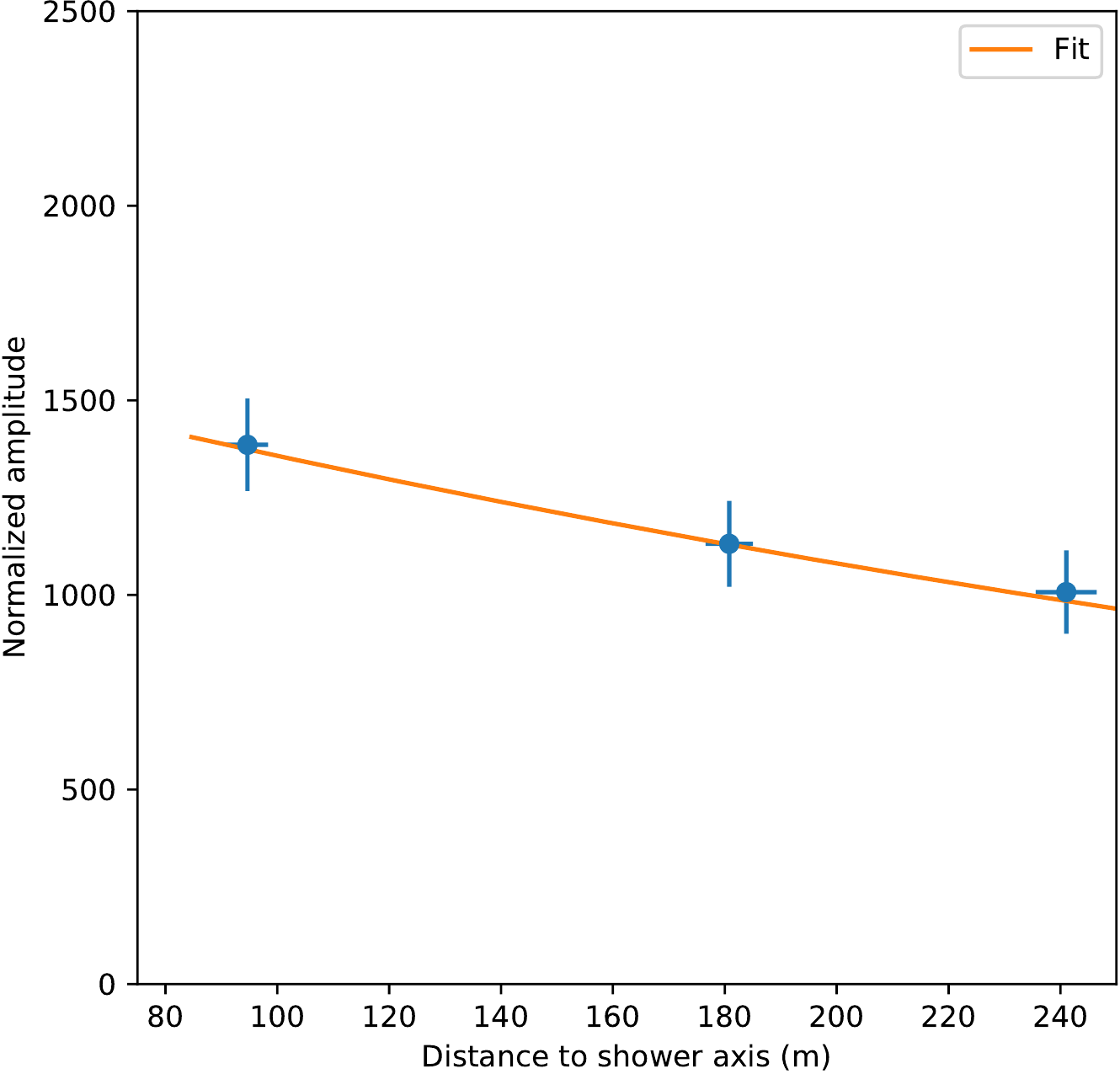}
\caption{\textit{Left:} The distribution of synthesised signals, which passed adaptive LDF processing.
Error bars include uncertainties introduced by synthesis procedure.
\textit{Right:} Average LDF for all detected events.
Orange curve indicated fit with exponential function, slope of which is used for the energy reconstruction.}
\label{fig:aldf}
\end{figure}

Left panel of Fig.~\ref{fig:aldf} shows the synthesised signals after processing with adaptive LDF algorithm.
One can note that adaptive LDF remove significant fraction of detected events, since synthesised signals do not reach sufficient level of signal-to-noise ratio (SNR).
The latest step is the energy reconstruction.
We have already developed algorithms for single antenna energy reconstruction~\cite{Tunka-Rex:2016gcn}, which uses parametrization based on average LDF.
We readjusted its parameters using LDF slope fitted from reconstructed events (see right panel of Fig.~\ref{fig:aldf}) and normalization from Tunka-Rex software corresponding to the actual calibration.
This way, primary energy $E_\mathrm{pr}$ is reconstructed using following:
\begin{equation}
E_\mathrm{pr} = \kappa E_\mathrm{ALDF} \exp\left[(d_\mathrm{ALDF}-d_0)\eta\right] / \sin\alpha\,,
\end{equation}
where $\kappa = 868\,\,\mathrm{EeV/(V/m)}$, $d_0 = 180$\,m, fitted $\eta = 0.00228$ and $\alpha$ is geomagnetic angle (angle between shower axis and magnetic field of the Earth).

\begin{figure}[h!]
\centering
\includegraphics[height=0.49\linewidth]{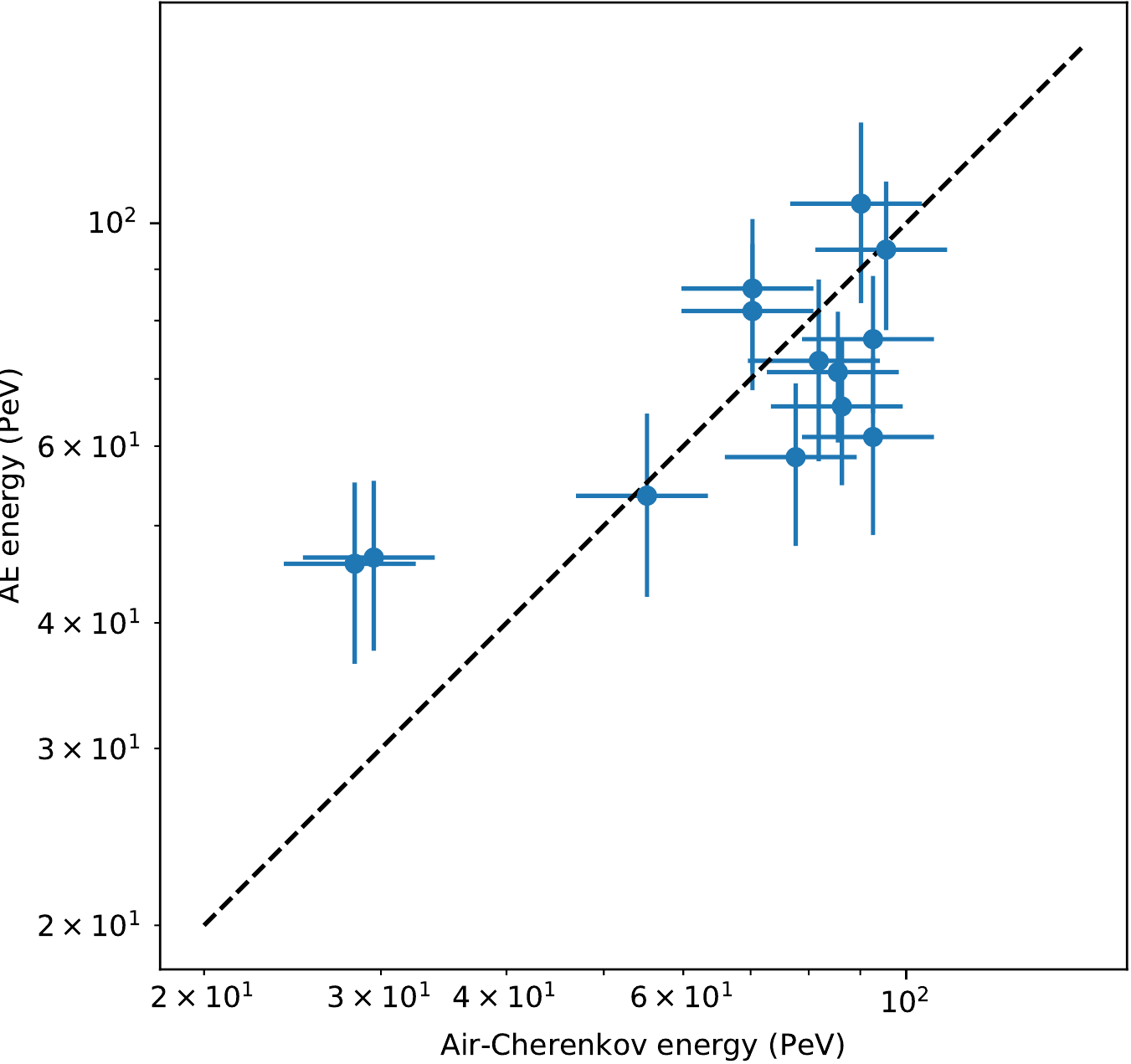}~~~\includegraphics[height=0.49\linewidth]{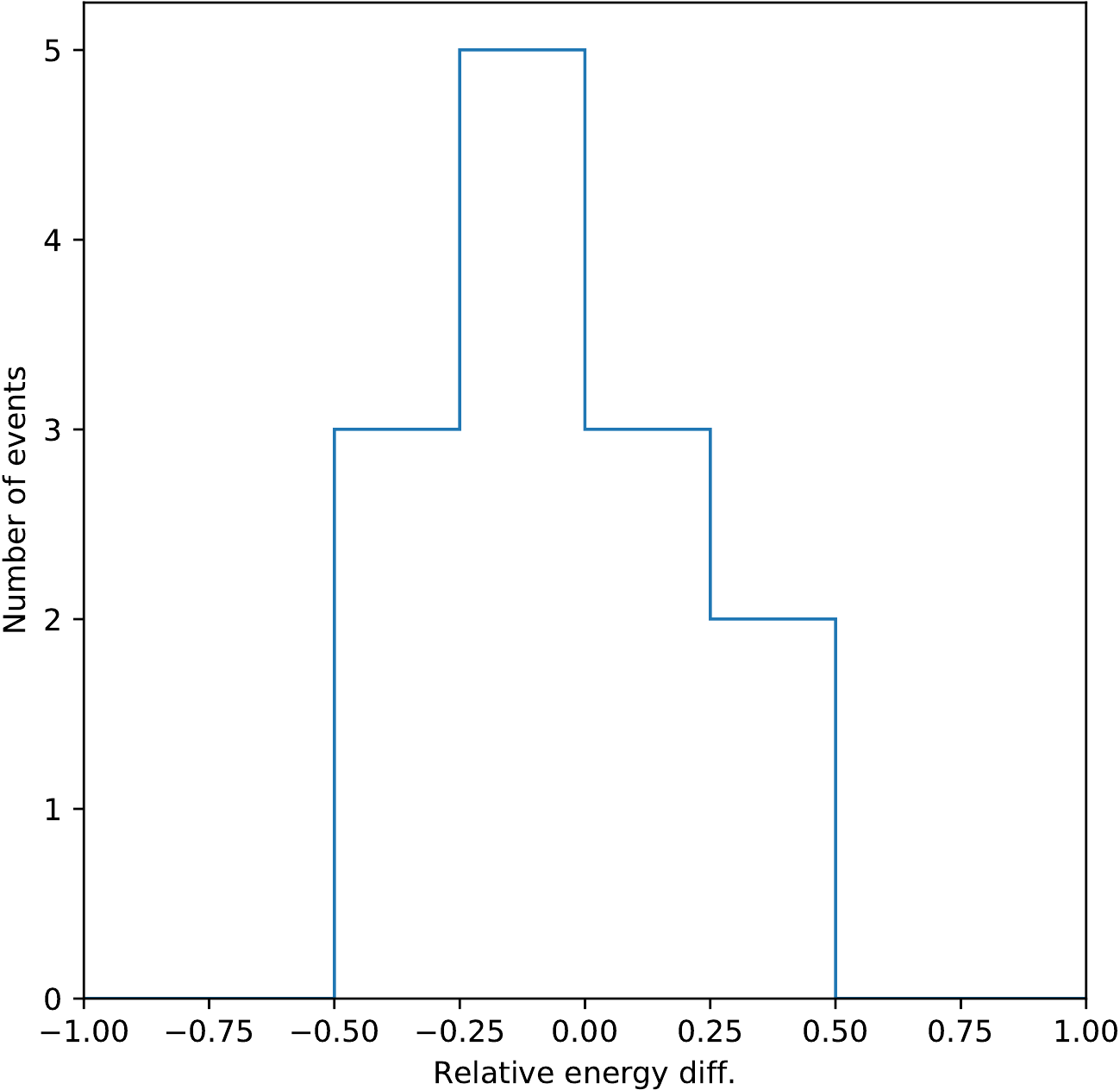}
\caption{Performance of energy reconstruction obtained for sub-threshold events.
\textit{Left:} cross-check with Tunka-133 reconstruction for the same events.
\textit{Right:} deviation between Tunka-133 and Tunka-Rex reconstruction showing the Tunka-Rex resolution of 26\%.}
\label{fig:energy}
\end{figure}

Fig.~\ref{fig:energy} shows the performance of the method for the energy reconstruction.
One can see that we have successfully reconstructed energy and arrival directions for two events with energies of 30\,PeV which is the record value for the Tunka-Rex.

\section{Conclusion}
We have reconstructed real data measured by the cosmic-ray radio detector using a deep neural network for a first time.
The algorithms developed in this works allow us to reconstruct events not detectable with standard Tunka-Rex methods obtained high angular and energy resolution.
The chosen architecture (autoencoder) and training pipeline (COREAS traces mixed with real background) show the efficiency of our approach.
The developed methods show high performance for the detection of the signal, which drives us to propose the following application of this technique:
\begin{itemize}
\item Sub-threshold detection and reconstruction of the signals at the tails of the LDF with low SNR is accessible with combination of autoencoder and adaptive LDF.
As shown in earlier studies~\cite{Kostunin:2015taa,Tunka-Rex:2016rkn}, the slope of the LDF characterized by its tails is sensitive to the position of shower maximum, which is a key for the identification of primary particle.
\item Trigger based on the compressed autoencoder integrated on FPGA~\cite{Duarte:2018ite} is a perspective technique for the implementation of the selt-triggered detector for the next generation of digital radio arrays~\cite{xie,selftrigger}.
We plan to test this approach in the nearest future.
\end{itemize}
The third-party code based on our software~\cite{abdul} has shown its efficiency for the broader frequency range, what indicates the universality of the approach.
Our software is published under free license\footnote{\url{https://gitlab.ikp.kit.edu/tunkarex/denoiser}} and we plan its further improvement.
One can find the example of usage of this software in JupyterHub at IAP KIT\footnote{\texttt{tutorials/jbr/trvo/trvo\_icrc.ipynb} at \url{https://jupyter.iap.kit.edu}}.
Further read can be also found in Refs.~\cite{Shipilov:2018wph, Bezyazeekov:2019jbe, Bezyazeekov:2020qqi}.
The autoeconder was also included in educational project at the first workshop of Mathematical center in Akademgorodok\footnote{\url{https://english.nsu.ru/mca/media/news/2973965/}}. 

\section*{Acknowledgements}
The authors would like to express gratitude to the colleagues from KCDC team.
The development and testing of the software was supported by the state contract with Institute of Thermophysics SB RAS.

\bibliographystyle{JHEP}
\bibliography{references}

\clearpage
\section*{Full Authors List: \Coll\ Collaboration}
\scriptsize
\noindent
P.~Bezyazeekov$^{1}$,
N.~Budnev$^{1}$,
O.~Fedorov$^{1}$,
O.~Gress$^{1}$,
O.~Grishin$^{1}$,
A.~Haungs$^{2}$,
T.~Huege$^{2,3}$,
Y.~Kazarina$^{1}$,
M.~Kleifges$^{4}$,
E.~Korosteleva$^{5}$,
D.~Kostunin$^{6}$,
L.~Kuzmichev$^{5}$,
V.~Lenok$^{2}$,
N.~Lubsandorzhiev$^{5}$,
S.~Malakhov$^{1}$,
T.~Marshalkina$^{1}$,
R.~Monkhoev$^{1}$,
E.~Osipova$^{5}$,
A.~Pakhorukov$^{1}$,
L.~Pankov$^{1}$,
V.~Prosin$^{5}$,
F.~G.~Schr\"oder$^{2,7}$
D.~Shipilov$^{8}$ and
A.~Zagorodnikov$^{1}$
~\\
~\\
\noindent
$^{1}$Applied Physics Institute ISU, Irkutsk, 664020 Russia\\
$^{2}$Karlsruhe Institute of Technology, Institute for Astroparticle Physics, D-76021 Karlsruhe, Germany\\
$^{3}$Astrophysical Institute, Vrije Universiteit Brussel, Pleinlaan 2, 1050 Brussels, Belgium\\
$^{4}$Institut f\"ur Prozessdatenverarbeitung und Elektronik, Karlsruhe Institute of Technology (KIT), Karlsruhe, 76021 Germany\\
$^{5}$Skobeltsyn Institute of Nuclear Physics MSU, Moscow, 119991 Russia\\
$^{6}$DESY, Zeuthen, 15738 Germany\\
$^{7}$Bartol Research Institute, Department of Physics and Astronomy, University of Delaware, Newark, DE, 19716, USA\\
$^{8}$X5 Retail Group, Moscow, 119049 Russia
\end{document}